# Detection of LDDoS Attacks Based on TCP Connection Parameters

Michael Siracusano
School of Computing, Electronics and Mathematics
Faculty of Science and Engineering
Plymouth University, UK
michael.siracusano@postgrad.
.plymouth.ac.uk

Stavros Shiaeles
School of Computing, Electronics and Mathematics
Faculty of Science and Engineering
Plymouth University, UK
stavros.shiaeles@plymouth.ac.uk

Bogdan Ghita
School of Computing, Electronics and Mathematics
Faculty of Science and Engineering
Plymouth University, UK
bogdan.ghita@plymouth.ac.uk

*Abstract*— Low-rate application layer distributed denial of service (LDDoS) attacks are both powerful and stealthy. They force vulnerable webservers to open all available connections to the adversary, denying resources to real users. Mitigation advice focuses on solutions that potentially degrade quality of service for legitimate connections. Furthermore, without accurate detection mechanisms, distributed attacks can bypass these defences. A methodology for detection of LDDoS attacks, based on characteristics of malicious TCP flows, is proposed within this paper. Research will be conducted using combinations of two datasets: one generated from a simulated network, the other from the publically available CIC DoS dataset. Both contain the attacks slowread, slowheaders and slowbody, alongside legitimate web browsing. TCP flow features are extracted from all connections. Experimentation was carried out using six supervised AI algorithms to categorise attack from legitimate flows. Decision trees and k-NN accurately classified up to 99.99% of flows, with exceptionally low false positive and false negative rates, demonstrating the potential of AI in LDDoS detection.

## I. Introduction

In an increasingly connected digital world, reliable and secure access to our online resources has never been more important. 60% of people in the UK are using e-banking, doubling in just nine years [17] and 1.61 billion of the world's population shop online [21]. But with reliance on these platforms comes a serious threat from malicious attackers seeking to degrade and deny access to these services.

Application layer attacks are a relative newcomer to the hacker's toolkit with the frequency and potency set to increase in the coming years [8]. Many of these attacks exploit weaknesses within the HTTP protocol. A subset of these attacks are known as low-and-slow, or low-rate DoS (LDoS). They do not require vast swathes of traffic to overload services and can go undetected with static fingerprints.

Research and mitigation advice regarding LDoS attacks are primarily focussed on altering the configuration of the webserver, rather than detection of attack streams. This can reduce the quality of service for legitimate users.

The aim of this study is to detect application layer LDDoS attacks through TCP flow metadata analysis, from a distributed attack. The specific application layer attacks that will be categorised will be: slowread, slowheaders and slowpost. All flows will be run against six supervised machine and deep learning algorithms for binary classification in to attack or legitimate flows. Analysis will highlight the best algorithms and features to be used for detection of LDDoS attacks..

## II. Related Works

LDDoS attacks are LDoS attacks that are launched from hosts distributed on the Internet. A low-rate DDoS (LDDoS) attack has significant ability to conceal its traffic because of its similarity with normal traffic Since LDoS was initially proposed in [8], a series of variants of LDoS attacks have been discussed, including: Reduction of Quality (RoQ) attacks [4] that exploit the performance vulnerability during a system's adaptation process and LDoS attacks targeting application servers (LoRDAS attacks) [9].

Whilst there has been much research in DoS/DDoS, LDoS/LDDOS attacks, there are some common weaknesses throughout literature that should be addressed, especially related to the quality of the data analysed. For instance, many studies used far too little and outdated data, up to two decades old, [2], [11], [1]. This will miss many of the modern attack vectors that utilise LDoS capabilities and reduce the ability to generalise conclusions to modern systems. Furthermore, mitigation advice based on reducing the number of connections allowed from a source IP address have the potential to reduce the QoS for users [12], this is a weakness that must be solved if effective mitigation of LDDoS attacks is to be employed.

A significant amount of the literature regarding LDoS and LDDoS attack mitigation have not utilised recent advances in machine learning and deep neural networks to its full extent. They have focussed on too few traffic features, [5], [10], [21], [12], [13]. An optimal number of features will reduce the quantity of false positives, whilst maintaining relatively low computational overheads. Artificial intelligence may also create an adaptive model, capable of understanding the standard traffic flows to a web server. This can be achieved through understanding the features of the TCP flow that may highlight and attack stream.

Through analysis of the best feature set and testing against various algorithms, it may be possible to find a solution that can detect the presence of LDoS and LDDoS attacks from TCP flows. This methodology has not been employed in literature analysed. It will produce a detection mechanism that will highlight LDDoS TCP flows, before they cause a DoS attack.

The proposed methodology seeks to overcome criticisms of the studies mentioned by improving the test datasets, diversity of features and number of algorithms used to ascertain the best combination to generate high accuracy.

## III. PROPOSED METHOD

In the section below, we will analysed the proposed methodology followed in steps

### A. Concept

As identified in the previous section, existing methods have had limited success in dealing with LDDoS, either due to generating excessive false positives and denying access to legitimate users or, to provide robustness, monitoring traffic from the volume perspective and producing false negatives. While it is very easy to note the limitations, identifying a solution to overcome them is a rather complex issue, as the chosen method should not rely on timing-related patterns, yet be able to discriminate attack traffic.

One possible approach is to consider the use of TCP connection characteristics to be used as discriminating features. From a functional perspective, the connection characteristics are indicative of the client behaviour. While for a typical client the main objective is to download the data and close the connection, an LDDoS connection is more likely to keep the connection open and occupy server resources in the process. In the case of slowread, slowheaders and slowpost, rather than terminate the connection, the attack connections aim to exhaust server resources through all transport-related alternatives – read or send data at a very low rate, maintain the connection open for as long as possible, and collectively, occupy server memory until legitimate clients are denied access. The connections associated with such an attack would not be detected by either rate- or volume-based features; at best, they may be identified as a poorly-connected client in terms of performance. However, looking at TCP connection parameters, with client-related parameters in particular, would indicate that the client is indeed aiming to stall the connection and drain server resources.

In this context, this paper investigates the use of TCP connection parameters as a discriminator for identifying LDDoS attacks. The method is aiming to differentiate the stalling behaviour of attack flows versus normal, legitimate connections, by separating network traffic into TCP connections, extracting the associated TCP parameters into a connections dataset, and passing the dataset through an AI-based analysis.

### B. Pre-processing

The raw traffic must be pre-processed in order to produce a dataset of TCP parameters dataset to be used as input for the method. The analysis must be robust in order to be able to cope with relatively large data rates, but also sufficiently complex to extract all the relevant TCP parameters. While a bespoke solution is likely to be the more accurate alternative, the TCPtrace analysis tool can deliver a comprehensive set of parameters in an automated fashion, including a total of 142 characteristics associated with a connection. While a significant proportion of these parameters are likely to be irrelevant for the LDDoS detection, particularly the ones that relate to the network performance, such as RTT or retransmissions, or connection control (SYN and FIN packet count), TCPtrace also extracts the necessary endpoint-based parameters, such as receiver-advertised window or initial congestion window, or stalling-related parameters, such as idle time.

In order to simplify the decision process, the full set may be processed through the chosen classification methods, which can subsequently determine the optimal subset to be used as preferred features.

### C. Classifier

The behaviour of a TCP connection is rather complex, with both endpoint- and network-related parameters influencing the perceived performance of a data transfer and the resources demand that it places on the server. Given this complexity, together with the input data, the classifier plays a critical role in the process, as its ability to discriminate between legitimate and attack connections depends on the relationship between parameters for both legitimate and attack traffic. To ensure a comprehensive evaluation of the TCP parameters feature set, the preferred option for the study was to process the resulting dataset through several classification methods, including Logistic Regression, k-NN, SVM, Decision Trees, Random Forest and Deep Neural Network. The results of the six classifiers were then compared in terms of accuracy, confidence matrix, false acceptance and rejection rates, and evaluation speed.

## IV. EVALUATION

The section below is divided into two sub-sections for better understanding how the evaluation was conducted.

### A. Environment and dataset

As indicated in the review section, prior studies used various attack datasets, but they all shared the same limitation, as were not necessarily focused on LDDoS attacks. For this study, the preferred alternative was to produce a bespoke set, to include a mix of genuine and attack traffic. A topology of 24 legitimate clients, 8 attackers, and one web server was created in GNS3. The creation of a simulated network capable of generating hundreds of thousands of TCP flows, enough for machine learning training and testing was fundamental to the success of this project. The simulation was created in GNS3, using containers for all network nodes. Figure 1 is a overview of the network topology.

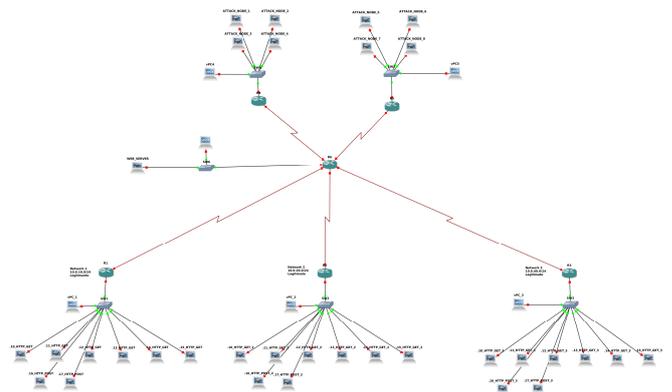

Figure 1 - Simulated network topography overview

The webserver was configured in a typical LAMP stack (Linux, Apache v2.4 HTTP server, MySQL database and PHP), using the default Apache settings and was set to run five websites. The legitimate activity was simulated and configured to randomly selected URLs stored on the webserver. It also simulated the HTTP POST method, by sending data to the MySQL database. In addition to traffic passing through at the unrestricted rate of 1Gbps, the scenario also included emulated slow connections by applying traffic shaping and throttling traffic to 11520 bps for some of the connections. The scenario included 84 hours of traffic generation, to include several types of attack, as summarised in Table 1. The resulting traffic was collected in pcap format using Wireshark on a network tap between the webserver and switch. The PCAP files were then processed using TCPtrace to extract the TCP connection parameters.

Table 1 - Summary of simulations performed and TCP Flow count

| Tool | Method | Traffic type | Duration | Node count | TCP Flow count |
|---|---|---|---|---|---|
| Siege | GET | Legitimate | 18 hours | 18 | 69969 |
| Siege | POST | Legitimate | 6 hours | 6 | 107299 |
| Siege | GET | Throttled | 24 hours | 18 | 10569 |
| Siege | POST | Throttled | 12 hours | 6 | 185471 |
| Slowread | GET | Attack | 8 hours | 8 | 80305 |
| Slowheaders | GET | Attack | 8 hours | 8 | 22312 |
| Slowbody | POST | Attack | 8 hours | 8 | 78325 |

In order to generalise the results and allow for comparison, the dataset was used in conjunction with the CIC Dataset, an application layer DoS dataset generated by the University of New Brunswick [6], filtered to include only web activity. Four datasets were produced, as listed in Table 2 below, by normalising, labelling and merging the simulated traffic and CIC dataset.

Table 2 – Test datasets

| Dataset | CIC Attack Flows | CIC Legitimate Flows | Simulated Attack Flows | Simulated Legitimate Flows | Total TCP Flows |
|---|---|---|---|---|---|
| 1 | 9311 | 9311 | 9305 | 9305 | 37233 |
| 2 | 9355 | 9355 | 0 | 0 | 18710 |
| 3 | 0 | 0 | 34586 | 34758 | 69344 |
| 4 | 0 | 66472 | 180942 | 373308 | 620722 |

Each of the four datasets has a slightly different purpose. Dataset 1 represents a balanced environment, with an equal amount of legitimate and attack simulated flows. Dataset 2 and 3 are to determine whether using a simulated environment has an impact on the accuracy of the estimation, as opposed to capturing traffic from a real network. Finally, dataset 4 evaluates the ability of the classification algorithms to generalise across a mix of simulated and real traffic by combining the legitimate traffic in the CIC dataset with the traffic from the simulations run.

*B. Processing*

Data processing involved three stages: feature selection, parameter tuning, and classification. Feature selection was achieved through Recursive Feature Elimination with 10-fold Cross Validation (RFECV) using SVM. The analysis indicated that the accuracy does not improve when increasing the number of features beyond 20 features, with the same subset making a significant impact on detection. To support the decision, some of the features appeared to be heavily correlated, as shown in Figure 2; the correlation analysis was run against dataset 1.

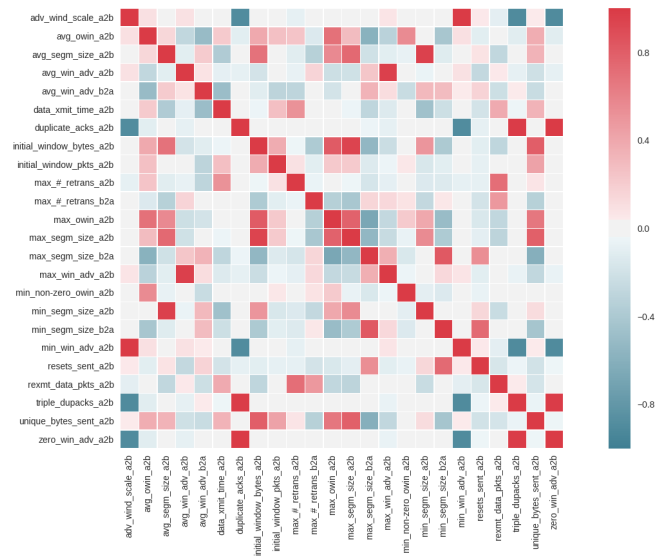

Figure 2 - Features correlation analysis

The automated RFECV analysis was then followed by a review of the parameters to determine correlated features. Following the two-stage selection, the dataset was filtered to include only the features giving the highest accuracy for LDDoS detection, as listed in Table 3.

Table 3 - Feature selection with descriptions

| Feature | Description |
|---|---|
| avg_win_adv_a2b | Average window advertisement. |
| data_xmit_time_a2b | Data transmit time. |
| max_win_adv_a2b | Maximum window advertisement. |
| throughput_a2b | Average throughput. |
| max_owin_a2b | Maximum outstanding unacknowledged data (bytes) |
| resets_sent_a2b | Number of reset packets sent. |
| avg_owin_a2b | Average outstanding unacknowledged data (bytes) |
| max_#_retrans_a2b | Retransmitted bytes, client to server. |
| min_segm_size_a2b | Minimum segment size. |
| initial_window_bytes_a2b | Total number of bytes sent in the initial window. |
| idletime_max_a2b | Maximum idle time between consecutive packets |
| idletime_max_b2a | Maximum idle time between consecutive packets |
| triple_dupacks_a2b | Triple duplicate ACKs sent from client to sever. |
| unique_bytes_sent_a2b | Total number of unique bytes sent. |

Six classifiers were chosen to determine the effectiveness of the TCP connection parameters in LDDoS detection: Logistic Regression (LG), k-NN (KNN), SVM, Decision Trees (DT), Random Forest (RF) and Deep Neural Network

(DNN). The aim behind the wide range of classifiers was to determine whether the choice of classifier does have a significant impact on the detection or the accuracy would be mainly due to the selected set of features. The analysis was performed using the TensorFlow machine learning framework [18] which allows automating the workflow data processing.

The classification stage involved running all six algorithms over the pre-processed input datasets. The output metrics for analysis were as follows: accuracy values from 10-Fold cross-validation, false positive rate (FPR), false negative rate (FNR). Besides accuracy, the algorithms were also assessed in terms of efficiency by monitoring the time it took to process to classify the data. In terms of training and testing, datasets 1-3 had a 1/2 train/test ratio and dataset 4 had a 10/90 test/train split.

## V. RESULTS

The table 4-7 below shows the results from the classification algorithms. We can see that k-NN and decision trees appeared to perform better, but all methods led to very high accuracy.

Table 4 - Results

| Dataset | | LG | KNN | SVC | DT | RF | DNN |
|---|---|---|---|---|---|---|---|
| 1 | Accuracy | 95.85 | 99.81 | 97.39 | 99.87 | 99.07 | 97.06 |
| | FPR | 7.76 | 0.08 | 4.09 | 0 | 0.58 | N/A |
| | FNR | 1.09 | 0.19 | 0.44 | 0.03 | 4.79 | N/A |
| | Eval time [s] | 0.023 | 0.558 | 1.87 | 0.019 | 1.842 | 471.9 |
| 2 | Accuracy | 94.82 | 99.96 | 99.96 | 99.96 | 99.93 | 99.96 |
| | FPR | 1.8 | 0 | 0 | 0 | 0.16 | N/A |
| | FNR | 9.65 | 0.03 | 0 | 0 | 0.13 | N/A |
| | Eval time [s] | 0.017 | 0.221 | 0.092 | 0.013 | 1.261 | 305.7 |
| 3 | Accuracy | 98.75 | 99.86 | 99.77 | 99.92 | 99.41 | 99.47 |
| | FPR | 1.77 | 0.06 | 0.26 | 0 | 0.56 | N/A |
| | FNR | 0.62 | 0.06 | 0.11 | 0.01 | 0.57 | N/A |
| | Eval time [s] | 0.034 | 1.096 | 1.727 | 0.029 | 2.951 | 510.8 |
| 4 | Accuracy | 92.75 | 99.9 | 99.35 | 99.92 | 99.41 | 99.47 |
| | FPR | 21.63 | 0.14 | 0.91 | 0 | 1.31 | N/A |
| | FNR | 1.3 | 0.03 | 0.54 | 0 | 0.14 | N/A |
| | Eval time [s] | 0.51 | 61.21 | 463.38 | 0.505 | 581.84 | 5203 |

Dataset 1 was used to observe how an equal mixture of legitimate and attack traffic would impact the classification. The results indicated excellent performance for k-NN, decision trees and random forests; given the results are based on 10-fold cross validation, the results do not suffer from overfitting. The exceptionally low FPR and FNR rates of the decision trees and k-NN indicate a value low enough to use within an IDS or IPS and demonstrate the potential of automated IP blocking for LDDoS prevention, as they would be unlikely to adversely affect a substantial quantity of legitimate flows. For decision trees, only two flows out of 6235 were inaccurately classed as legitimate with also the fastest evaluation time of all the classifiers of 0.019s for over 12000 TCP flows, using 14 features. This speed further highlights the potential for real-time flow categorisation. At the other end of the spectrum, the inaccuracy of both logistic regression and SVM was due to their bias towards categorisation of attack traffic, leading to an unacceptably high FPRs of 7.78% and 4.09% and being almost 2% less accurate than the better performing classifiers. Aside from the performed parameter tuning, it is possible to slightly improve on the results and lead to a more balanced FPR and FNR split. However, the FNR rates are high, even in this biased setting relative to the more accurate classifiers, which indicates they are unlikely to be more accurate than decision trees or k-NN. Finally, the deep neural networks performed disappointingly compared to the more successful classifiers, achieving the second lowest accuracy of 97.061%. Similar to the parameter tuning, results could be marginally improved with a larger number of epochs, but it is unlikely that the increase would be statistically significant.

Dataset 2 was based on the CIC data. As it can be seen in the results, all classifiers performed excellent, with both SVM and decision trees having no false positives or false negatives, and k-NN only having one malicious TCP stream misclassified as legitimate. The exception was logistic regression, which had a high false positive rate of over 9% and its accuracy was more than 5% below the other classifiers. An additional observation that can be made based on these results is that the high categorisation is possibly indicative of the lack of artificially generated slow traffic; this was the purpose of the throttled traffic within the simulated network activity. Given this, it is possible that some of the detection is based on the algorithms categorising slow and fast TCP activity, as the attack traffic emulates very slow TCP connections.

Dataset 3 was a single dataset experiment consisting of a balanced mix of over 70K TCP flows from the simulated network. The accuracy of k-NN, SVM and decision tress was exceptional, with decision trees having wrongly classified only one flow in the test data subset. The high accuracy for k-fold cross validation indicates a well-fitting model, capable of performing well on new input data from this network. The deep learning algorithms also performed very well with 99.962% accuracy; the learning process plateaued after 50 epochs, so it may be worth exploring as part of future research whether a lower learning rate could further improve the results. It is worth noting that, as part of this simulation, the traffic was throttled to simulate legitimate TCP connections running over slow links. This is likely to have caused the a high FNR for the logistic regression method, but it did not affect the other algorithms, which were all successful in classifying the activity.

Dataset 4 aimed to observe the ability of the algorithm to classify a highly unbalanced dataset, which includes a significant amount of legitimate traffic from both the CIC dataset and simulated network. Despite the unbalanced nature of the dataset the accuracy for all algorithms, except logistic regression, were very high. Decision trees is the clear best with a FPR and FNR of 0.001% and 0.004%, respectively. This represents only misclassification of only 11 TCP flows out of over half a million flows. k-NN performed well also, with FPR and FNR of 0.031% and 0.14%, respectively. The main advantage over decision trees compared to k-NN is not simply the accuracy, however. The evaluation time for k-NN is 61.21s while decision trees is only 0.505s. In an IDS/IPS this evaluation time is vital to avoid processing overload and to cope evaluating vast quantities of TCP connections to a busy webserver. This is where the speed and accuracy of decision trees will show the true benefit as each flow will take a millisecond to categorise and potentially block. This is a significant step towards the mitigation of LDDoS attacks. Once again, logistic regression is the lowest performing classifier, with a very high FNR of 21%. This indicates bias and high sensitivity for the data and can be seen in the ROC curve gradients and the low precision score of 90.163%. Tweaking the classifier parameters could mitigate some of the inaccuracies. However, it is unlikely to perform better than decision trees with its near perfect score and slightly faster (by 0.07s) evaluation time. The accuracy of the deep neural network was relatively high at 98.954%. However, the graph indicates the gradient still converging, rather than plateauing. With additional epochs, this may have improved a few hundredths of a percent. However, the training and evaluation time for a dataset this size was high, being 86 minutes for evaluation compared to 0.505 seconds for decision tree evaluation. With a lack of powerful parallel processing capabilities running the tests for longer would not significantly alter the conclusions of this research. These results highlight how well this methodology can work across very large datasets, with relatively few training examples. The near perfect results from the decision trees once again shows how useful this algorithm is for classification of attack and legitimate streams.

## VI. Discussion

The primary aim of the research was to ascertain if machine learning, using TCP performances as features, could be employed to detect malicious LDDoS attacks amongst legitimate activity, utilising artificial intelligence and TCP flow metadata. The results of the validation tests indicate that accurate categorisation can indeed be achieved. Both k-NN and decision trees accurately detect the activity with such low FPR and FNR that once the model is trained, they could be used to automatically block the attacking flows. significantly reducing the danger of these attacks. The techniques employed within this project categorise the TCP flows independent of the source IP address, thereby making equally effective against both single source and distributed attacks.

Through analysis of these results clearly machine and deep learning algorithms performed well across all the datasets. The deep learning model achieved an accuracy of 97-99.9%. However, the complexity of neural networks did not outperform the lightweight machine learning algorithms. The relatively significantly longer time taken to train and test these networks is an unnecessary drain on computational resources, that does not lead to an improvement over faster and more lightweight algorithms such as decision trees and k-NN. It is possible that will lower learning rates, more data and thousands more epochs accuracy could be improved. This is unlikely to be a proportional gain relative to the extra computational overheads, especially in the presence of other fast and accurate algorithms.

The extra computational time require for random forests led to less accurate results than the decision trees in all the tests. This could be due to the time and effort taken to pick relevant and accurate features in stage 1 of the evaluation. Logistic regression was the least accurate throughout, with Dataset 1 having almost double the FPR of the second least successful algorithm. It is possible that there were issues with the parameter selection of the classifier, causing excessive bias and sensitivity throughout the classification of the activity. The most accurate algorithms throughout were k-NN and decision trees. They were capable of extremely low FPR and FNR across all datasets and tests. The evaluation time per flow and consistently high accuracy across all tests for decision trees makes it the clear favourite algorithm for binary classification of LDDoS activity. The speed and the extremely low FPR and FNR demonstrate that this could be used for automated detection and mitigation of malicious TCP flows real-time in an IPS system.

Dataset 4 had a dual purpose – to determine the capacity of the method to generalise when presented with other traffic, but also to determine its robustness, as the test/train ratio was 10/90, with 2.5 times more legitimate activity. Even with this data, all models bar logistic regression and DNN achieved over 99.3% k-fold validation precision and with low FPR and FNR. It was anticipated that decision trees were likely to be the most successful, which was confirmed by experimental validation. The lack of accuracy with the logistic regression could highlight the data not being linearly separable, as this tends to be more accuracy for classifying data of that sort. The SVM RBF kernel outperformed the linear kernel, which supports this conclusion.

The main limitation of this research is that fact it was carried out entirely on two datasets, based on simulated attacks, which may be perceived as slightly biased. Lack of parallel GPU processing capabilities consequently led to lower epochs and higher learning rate than would have been optimal for achieving the best accuracy level. Volume of data is also an important aspect of machine learning for training, testing and validation. More data could result that models could be created that could be applicable to a variety of server set ups and content..

## VII. CONCLUSION AND FUTURE WORK

LDDoS attacks are well known for their stealth, with static snort signatures unable defend and the traditional volumetric DoS and DDoS attack mitigation strategies being redundant. Most mitigation advice focuses on the necessity of limiting the number of connections per IP address to the webserver. This will be ineffective for a distributed attack, which would still require far less zombie nodes than a standard volumetric attack.

The primary aim of the research was to ascertain if machine learning could be used to detect malicious LDDoS attacks amongst legitimate activity, utilising artificial intelligence and TCP flow metadata. Following an evaluation of a number of algorithms over a mix of simulated and real data, it was demonstrated that stealthy application layer distributed LDoS attacks can be accurately categorised from legitimate traffic using features associated with their TCP flows. A range of AI algorithms can use these features to accurately predict the presence of attack streams; given the evaluated algorithms, the most accurate ones appeared to be decision trees and k-NN.

The accuracy achieved within the experiments is comparable or an improvement on much of the research analysed within the literature review. [15] for a slow read application layer DoS attack achieved accuracy of 99.37% with their random forest classifiers. This experiment achieved between 99.41% and 99.94%. Whilst only a slight increase, in a real word environment this can mean the difference between a successful attack and one that is mitigated. [4] in their study on IP flows of botnets achieved 97% accuracy for detection of Citadel and 86% for Zeus using decision trees and Naïve Bayes. The decision trees within across all four experiments were higher, with k-fold cross validated figures between 99.88% and 99.97%. [14] trained neural networks to detect a variety of DoS attacks from packet header analysis and achieved and accuracy of 98%. A Probabilistic Neural Network trained to detect attacks by [17] achieved 97.89% accuracy. The DNN used within these experiments achieved between 97% and 99.9%.

Future work could involve using a wider range and larger volume of real-world data. Cloud computing platforms capable of network simulation, data storage, data analytics and machine learning can provide better data sources and processing capability. Lastly, integration within an aggregated SIEM solution could allow a richer feature set, including webserver metrics such as connection events, CPU load and memory usage. This would likely lower the chances of false positives outside of a simulated environment.

ACKNOWLEDGMENT

This work was supported by CYBER-TRUST project, which has received funding from the European Union's Horizon 2020 research and innovation programme under grant agreement no. 786698.